\documentclass[
 reprint,
 amsmath,amssymb,superscriptaddress,
 aps,
 prb,
]{revtex4-1}

\usepackage{graphicx}
\usepackage{dcolumn}
\usepackage{bm}

\begin{document}

\title{Ballistic transport of (001) GaAs 2D holes through a strain-induced \\ lateral superlattice}

\author{D.\ Kamburov}
\author{H.\ Shapourian}
\author{M.\ Shayegan}
\author{L.N.\ Pfeiffer}
\author{K.W.\ West}
\author{K.W.\ Baldwin}
\affiliation{
 Department of Electrical Engineering, Princeton University, Princeton, New Jersey 08544, USA}
\author{R.\ Winkler}
\affiliation{Department of Physical Chemistry, The University of the Basque Country, 48080 Bilbao, Spain}
\affiliation{IKERBASQUE, Basque Foundation for Science, 48011 Bilbao, Spain}

\date{\today}

\begin{abstract}
We report the observation of ballistic commensurability oscillations and positive magnetoresistance in a high-mobility, (001) GaAs two-dimensional hole system with a unidirectional, surface-strain-induced, periodic potential modulation. The positions of the resistivity minima agree well with the electrostatic commensurability condition. From an analysis of the amplitude of the oscillations we deduce a ballistic scattering time and an effective magnitude for the induced periodic potential seen by the two-dimensional holes.
\end{abstract}


\maketitle

Spatially modulated two-dimensional (2D) carrier systems exhibit magnetoresistance oscillations as a consequence of the commensurability between the size of the ballistic cyclotron orbits and the period of the modulation. \cite{ref01Weiss.EPL.8.179.1989,ref02Winkler.PRL.62.1177.1989,ref03Gerhardts.PRL.62.1173.1989,ref04Beenakker.PRL.62.2020.1989,ref05Beton.PRB.42.9229.1990,ref06Peeters.PRB.46.1992,ref06bMirlin.PRB.58.1998,ref07Lu.PRB.58.1138.1998,ref13Lu.PRL.81.1282.1998;Lu.PRB.60.13776.1999,ref08Smet.PRL.83.2620.1999,ref09Skuras.APL.70.871.1997,ref10Endo.PRB.62.16761.2000,ref11Endo.PRB.72.235303.2005,ref12Endo.JPSJ.74.1792.2005} These commensurability oscillations (COs) have proved to be a useful tool for studying transport properties as they lead to an interplay between the inherent parameters of the 2D carriers and those of the modulation potential. The positive magnetoresistance, typically seen at very low fields, and the amplitude of the COs are related to the magnitude of the modulation potential and the scattering times. The frequency of the COs probes the shape and size of the Fermi contours and the subband occupations. While most investigations of commensurability effects have been performed on GaAs 2D electron systems, \cite{ref01Weiss.EPL.8.179.1989,ref02Winkler.PRL.62.1177.1989,ref03Gerhardts.PRL.62.1173.1989,ref04Beenakker.PRL.62.2020.1989,ref05Beton.PRB.42.9229.1990,ref06Peeters.PRB.46.1992,ref06bMirlin.PRB.58.1998,ref07Lu.PRB.58.1138.1998,ref08Smet.PRL.83.2620.1999,ref09Skuras.APL.70.871.1997,ref10Endo.PRB.62.16761.2000,ref11Endo.PRB.72.235303.2005,ref12Endo.JPSJ.74.1792.2005} GaAs 2D hole systems (2DHSs) have been studied as well. \cite{ref13Lu.PRL.81.1282.1998;Lu.PRB.60.13776.1999,ref14footnote1} The 2DHS studies, however, have primarily focused on low-symmetry (113)$A$ GaAs samples because high-quality, C-doped (001) GaAs samples have not been available until recently. \cite{ref15footnote2}

Here we report a quantitative study of COs in high-mobility, C-doped (001) GaAs 2DHSs. To create a periodic, unidirectional potential modulation, we employ surface strain induced by stripes of negative electron-beam (e-beam) resist deposited on the sample surface. It produces a local 2DHS density modulation through the piezoelectric effect. \cite{ref09Skuras.APL.70.871.1997} The successful use of such a technique for inducing COs has been demonstrated in GaAs 2D electron systems. \cite{ref09Skuras.APL.70.871.1997,ref10Endo.PRB.62.16761.2000,ref11Endo.PRB.72.235303.2005,ref12Endo.JPSJ.74.1792.2005} Our measurements reveal resistance minima at the electrostatic commensurability condition: \cite{ref01Weiss.EPL.8.179.1989,ref02Winkler.PRL.62.1177.1989,ref03Gerhardts.PRL.62.1173.1989,ref04Beenakker.PRL.62.2020.1989,ref05Beton.PRB.42.9229.1990,ref06Peeters.PRB.46.1992,ref06bMirlin.PRB.58.1998,ref07Lu.PRB.58.1138.1998,ref13Lu.PRL.81.1282.1998;Lu.PRB.60.13776.1999,ref08Smet.PRL.83.2620.1999,ref09Skuras.APL.70.871.1997,ref10Endo.PRB.62.16761.2000,ref11Endo.PRB.72.235303.2005,ref12Endo.JPSJ.74.1792.2005}
\begin{equation}\frac{2R_C}{a}=i-\frac{1}{4}\;\;\;(i=1,2,3,...),
\label{eq:1}
\end{equation}
where $R_C$ is the cyclotron radius, $a$ is the period of the potential modulation, and $2R_C=2\hbar k_F/eB$ with $k_F=\sqrt{2\pi p}$ and $p$ being the Fermi wave vector and hole density, respectively. We present the dependence of the amplitude and positions of the COs on the potential modulation period and the 2DHS density. We also extract estimates of the magnitude of the potential modulation and the scattering time for the ballistic holes.

\begin{figure}[t]
\includegraphics[width=0.48\textwidth]{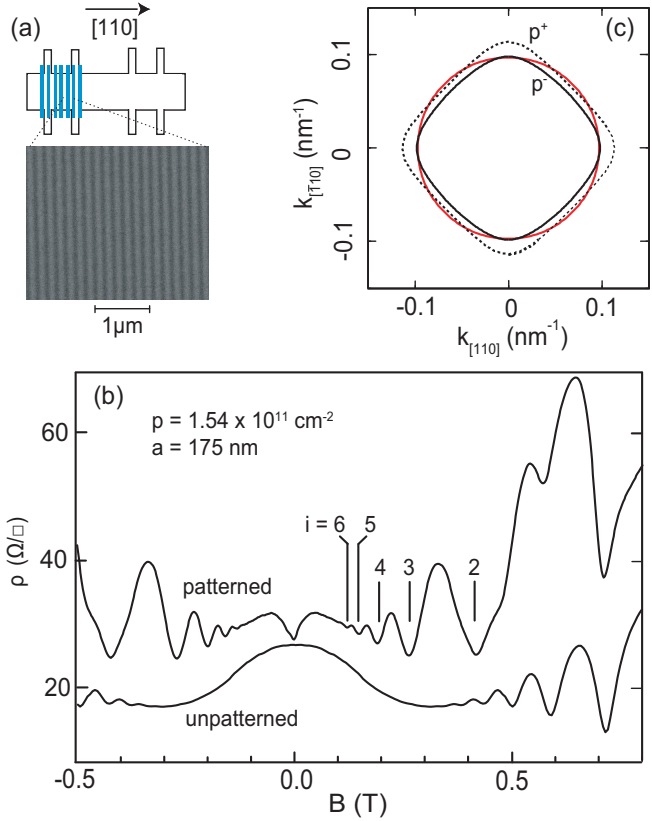}
\caption{\label{fig:Fig1} (color online) (a) Schematic of the Hall bar showing the patterned and unpatterned regions as well as a scanning electron microscope image of the negative e-beam resist pattern for the $a$ = 175 nm period sample. (b) Magnetotransport data from the patterned (upper trace) and unpatterned (lower trace) regions obtained by passing current through the Hall bar and measuring the voltage along each region. The vertical lines mark the expected positions of the resistance minima corresponding to index $i$ (see Eq.\;\eqref{eq:1}). (c) Self-consistently calculated Fermi contours $p^+$ and $p^-$ for this sample at $p=1.50 \times 10^{11}$ cm$^{-2}$ along with the spin-degenerate, circular Fermi contour (red), corresponding to $k_F=\sqrt{2 \pi p}$.}
\end{figure}

We prepared strain-induced lateral superlattice samples with different lattice periods from a 2DHS confined to a 175-\AA\--wide GaAs quantum well (QW) grown via molecular beam epitaxy on an undoped (001) GaAs substrate. The QW is located 131 nm below the surface and is surrounded on each side by 95-nm-thick undoped Al$_{0.24}$Ga$_{0.76}$As (spacer) and C $\delta$-doped layers. The as-grown 2DHS density at 300 mK is $p=1.55\times10^{11}$ cm$^{-2}$, and the mobility is $\mu=1.2\times10^{6}$ cm$^2$/Vs. The density was varied using an In back gate. Hall bars were made by wet etching past the dopants along the $[110]$ or $[\overline{1}10]$ directions. As illustrated in Fig.\;\ref{fig:Fig1}(a), half of each Hall bar was covered with a grating of high-resolution negative e-beam resist using e-beam lithography while the other half was kept unpatterned as a reference. The periods of the gratings were 100, 150, 175, 200, 250, and 300 nm. A scanning electron microscope image of a 175-nm-period resist grating is given in Fig.\;\ref{fig:Fig1}(a). The samples were measured in a pumped $^3$He refrigerator with a base temperature of $T\simeq$ 300 mK using lock-in techniques.

\begin{figure}[t]
\includegraphics[width=0.48\textwidth]{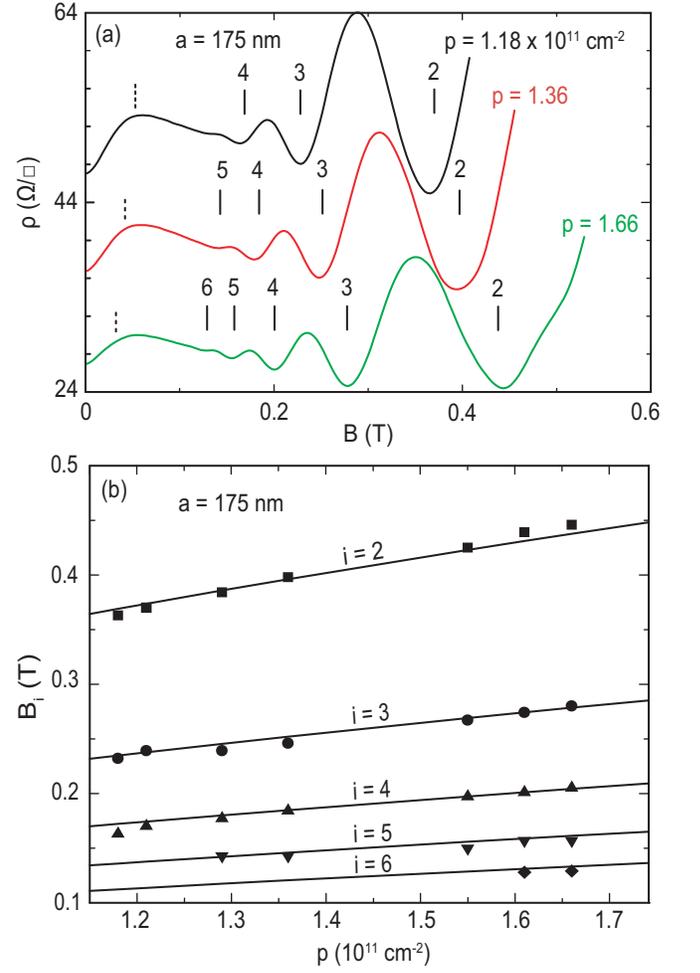}
\caption{\label{fig:Fig2} (color online) (a) Magnetotransport data at different densities for the $a=175$ nm sample. Vertical indexed lines mark the positions of the resistance minima based on spin-degenerate, circular Fermi contours. The dashed vertical lines show the position of the extinction field determined from the envelope fitting (see text). (b) Summary of the observed (symbols) and expected (lines) positions of the CO minima for the $a=175$ nm period sample.}
\end{figure}

Low-field magnetoresistance traces from the patterned and unpatterned regions of the $a=175$ nm sample are shown in Fig.\;\ref{fig:Fig1}(b). The trace from the unpatterned region shows Shubnikov-de Haas (SdH) oscillations starting around $B=0.4$ T. The modulated region's trace exhibits clear COs below 0.4 T. The expected positions of the COs minima, based on a circular, spin-degenerate Fermi contour, i.e. with $k_F=\sqrt{2  \pi p}$, are marked with vertical lines indexed according to Eq.\;\eqref{fig:Fig1}. Because of the mixing between the CO and SdH oscillations at $B>0.4$ T, the minimum corresponding to $i=1$ is difficult to distinguish and is not shown here. The trace from the patterned region also exhibits positive magnetoresistance at very small fields, signalling the formation of resonant orbits. \cite{ref04Beenakker.PRL.62.2020.1989,ref05Beton.PRB.42.9229.1990}

Figure \;\ref{fig:Fig1}(c) shows the results of self-consistent calculations of the Fermi contours for a 2DHS at $p=1.50\times 10^{11}$ cm$^{-2}$ confined to a 175-\AA\--wide (001) GaAs QW. The calculation is free of adjustable parameters and is based on the $8 \times 8$ Kane Hamiltonian. \cite{ref16Winkler.Springer.2003} It accounts for the nonparabolicity, spin-orbit interaction induced spin-splitting, and the anisotropy of the 2DHS band structure. There are two anisotropic Fermi contours, $p^+$ and $p^-$, one for each spin-subband. Because the Fermi wave vectors for $p^+$ and $p^-$ along the current direction are different, in principle, there should be two sets of COs with different frequencies. \cite{ref13Lu.PRL.81.1282.1998;Lu.PRB.60.13776.1999} However, our data exhibit only one set of COs, consistent with the red circular contour corresponding to $k_F=\sqrt{2 \pi p}$ shown in Fig.\;\ref{fig:Fig1}(c), and suggesting that we cannot resolve the two contours. \cite{ref17additionalfootnote}

The dependence of the COs on density for the $a=175$\;nm period sample is presented in Fig.\;\ref{fig:Fig2}. The density is varied from 1.18 to 1.66 $\times$ 10$^{11}$ cm$^{-2}$ using the back gate. Magnetoresistance traces are shown for three representative densities in Fig.\;\ref{fig:Fig2}(a). Figure\;\ref{fig:Fig2}(b) summarizes the evolution of all the observed CO minima along with their values according to Eq.\;\eqref{eq:1}, and assuming spin-degenerate, circular Fermi contours.

\begin{figure}[t]
\includegraphics[width=0.48\textwidth]{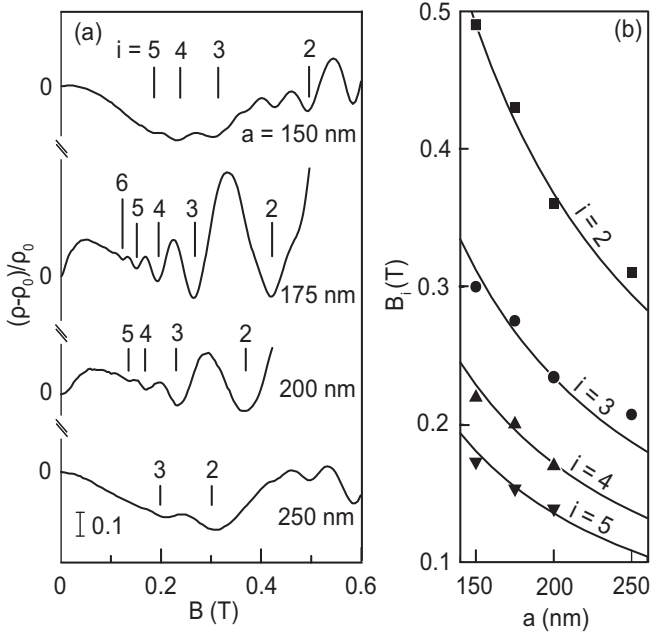}
\caption{\label{fig:Fig3} (a) Magnetoresistance data of samples with $p\simeq1.53\times$10$^{11}$ cm$^{-2}$ and different periods. The vertical axis shows the fractional change of the magnetoresistance with respect to the zero-field value. The traces are offset vertically without changing the scale. Indexed vertical lines mark the predicted positions of the COs minima. (b) Summary of the observed (symbols) and expected (lines) positions of the COs minima for different periods at $p$ $\simeq$ 1.53 $\times$ 10$^{11}$ cm$^{-2}$.}
\end{figure}

Figure\;\ref{fig:Fig3}(a) shows the low-field magnetoresistance of samples with different modulation periods and with $p\simeq1.53\times$10$^{11}$ cm$^{-2}$. The observed CO minima are close to the predicted positions for all periods. Figure\;\ref{fig:Fig3}(b) summarizes the evolution of the observed minima with modulation period at a fixed density of $p\simeq1.53\times$10$^{11}$ cm$^{-2}$. Data from the 100 nm and 300 nm period samples are not shown here as these samples do not exhibit any COs at all, and traces from their patterned and unpatterned regions are very similar. The positive low-field magnetoresistance appears to be even more sensitive to the modulation period than the COs themselves and is discernible only for the 175 nm and 200 nm period gratings.

The oscillatory part of the COs, $ \rho^{osc}$, can be expressed in terms of the ratio $ \eta=V_0/E_F$ of the potential modulation amplitude $V_0$ to the Fermi energy $E_F$ by treating the modulation as a perturbation. \cite{ref04Beenakker.PRL.62.2020.1989,ref05Beton.PRB.42.9229.1990,ref06Peeters.PRB.46.1992,ref06bMirlin.PRB.58.1998} The resulting asymptotic expression:
\begin{equation}\frac{ \rho^{osc}}{ \rho_{0}}=A\left( \frac{ \pi}{ \omega_c \tau_{CO}} \right)\frac{ \eta^2}{2} \frac{L}{a}  \mu B \sin\left(2 \pi\frac{2R_C}{a} \right)
\label{eq:2}
\end{equation}
includes the mean-free path $L=\hbar k_F  \mu /e$, where $ \mu$ is the mobility, and an additional damping factor $A(\pi / \omega_c \tau_{CO})$. \cite{ref10Endo.PRB.62.16761.2000} The damping factor, with $A(x)=x/\sinh(x)$, $ \omega_c$ the cyclotron frequency, and $ \tau_{CO}$ the CO lifetime, accounts for carriers that scatter before completing a cyclotron orbit. The temperature factor given in Ref.\citenum{ref10Endo.PRB.62.16761.2000} has been omitted here because its value is a constant nearly equal to unity for temperatures close to 300 mK. 

\begin{figure}[t]
\includegraphics[width=0.48\textwidth]{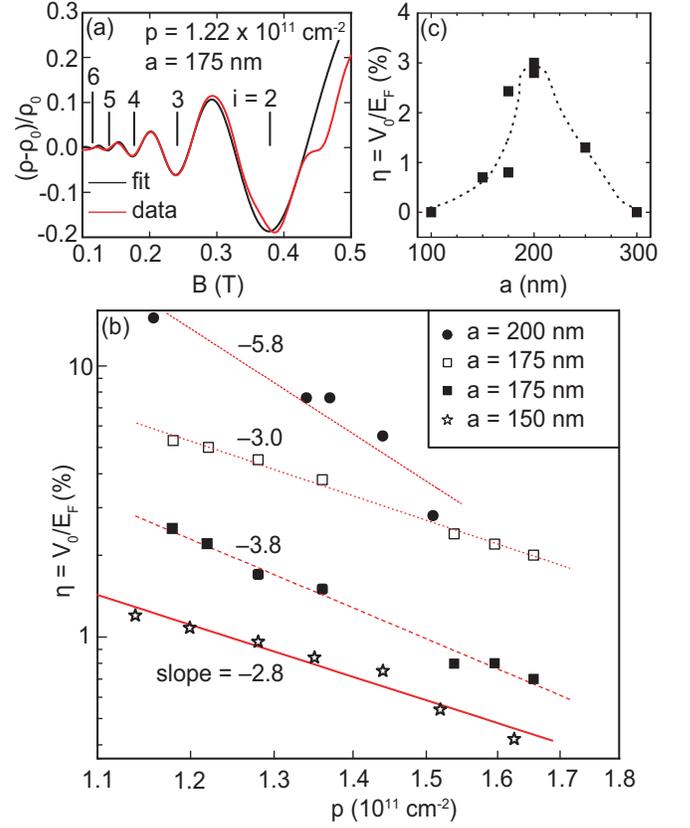}
\caption{\label{fig:Fig4} (color online) (a) Oscillatory part of the magnetoresistance data along with a fitted curve based on Eq.\;\eqref{eq:2}. (b) $ \eta$ from fits of magnetotransport data for samples with periods 150, 175, and 200 nm as a function of density in a log-log plot. (c) $ \eta$ as a function of period at fixed density $p \simeq 1.53 \times 10^{11}$ cm$^{-2}$. The dotted line is a guide to the eye.}
\end{figure}

We studied the dependence of the ratio $ \eta$ on the period of the modulation and the 2DHS density quantitatively by fitting the envelope of the COs using Eq.\;\eqref{eq:2}. The results are shown in Fig.\;\ref{fig:Fig4}. We fit the values of the maxima and minima for each magnetoresistance trace before the appearance of SdH oscillations independently with spline curves. We then use the difference between the two fits to obtain the background resistance which is subtracted from the magnetoresistance trace to yield its oscillatory part. The oscillatory part of a typical trace is plotted in Fig.\;\ref{fig:Fig4}(a) along with the fit using Eq.\;\eqref{eq:2}. The calculated positions of CO minima are given as well.

The fit shown in Fig.\;\ref{fig:Fig4}(a) gives a CO lifetime $\tau_{CO} \simeq 5$ ps. In this estimate we used the effective mass obtained from our band calculations, $m^*=0.2m_e$, where $m_e$ is the free-electron mass. The transport (mobility) lifetime obtained from the zero-field resistance is $ \tau_{ \mu}=145$ ps. The SdH (quantum) lifetime estimated from the $B$-dependence of the envelope of the SdH oscillations in the unpatterned region above $B=0.4$ T is $ \tau_{SdH} \simeq 1$ ps. The observation that $\tau_{CO} > \tau_{SdH}$ is somewhat surprising because, according to the available theory, \cite{ref06Peeters.PRB.46.1992} $\tau_{CO}$ and $\tau_{SdH}$ should have the same value. Our observation implies that COs are not as sensitive to (small-angle) scattering as SdH oscillations are. We note that $\tau_{CO} > \tau_{SdH}$ has also been reported for other 2D carrier systems.\cite{ref07Lu.PRB.58.1138.1998,ref13Lu.PRL.81.1282.1998;Lu.PRB.60.13776.1999} Additionally, the value of our deduced $\tau_{CO}$ is comparable to the CO lifetimes reported for (113)$A$ GaAs 2DHSs.\cite{ref13Lu.PRL.81.1282.1998;Lu.PRB.60.13776.1999}

Figure\;\ref{fig:Fig4}(b) summarizes the deduced ratio $ \eta$ as a function of density for several samples. The data for $a=175$ nm are from two Hall bars on the same sample. The log-log scale in Fig.\;\ref{fig:Fig4}(b) permits us to extract an approximate power-law for the dependence of $ \eta$ on the density. If we assume that the modulation amplitude $V_0$ does not depend on density, $ \eta$ is anticipated to have a $p^{-1}$ dependence coming from the linear relation between the Fermi energy $E_F$ and the hole density, $E_F \propto p$ . Similar to previous results for electrons, \cite{ref10Endo.PRB.62.16761.2000} however, the data for all samples (Fig.\;\ref{fig:Fig4}(b)) show that $ \eta$ decreases much faster with $p$. This observation, for both electrons and holes, likely stems from screening which leads to a reduction of the modulation potential amplitude with increasing density. 

Figure\;\ref{fig:Fig4}(c) is effectively a vertical cut at a fixed density of $p\simeq 1.53 \times $10$^{11}$ cm$^{-2}$. It includes additional data points from other samples. The dependence of $ \eta$ on period implies that $ \eta$ obtained from the envelope fits reaches a maximum when the period is 200 nm and goes to zero for periods $\leq$ 100 nm and $\geq$ 300 nm. In the $a=100$ nm sample, the modulation is likely smeared out because the period is smaller than the distance of the 2DHS from the surface. We do not know why the $a=300$ nm sample does not show COs. Based on the COs observed at large $i$ in samples with shorter modulation period, the mean-free-path should be long enough for the 2D holes in this sample to complete ballistic trajectories.

The CO envelope fits can also be used to estimate the positions of the expected extinction field $B_E$ marked by the vertical dashed lines in Fig.\;\ref{fig:Fig2}(a). \cite{ref05Beton.PRB.42.9229.1990} This is the field at which the Lorentz force balances the force of the electrostatic potential. When $B>B_E$, holes overcome the confinement potential created by the electrostatic modulation and resonant orbits are no longer formed, resulting in a decrease of the resistivity. The estimation of the position of $B_E$ is done using the relation between the extinction field and the ratio $ \eta$: \cite{ref10Endo.PRB.62.16761.2000, ref11Endo.PRB.72.235303.2005}
\begin{equation} B_E=\frac{V_0}{E_F}\left(\frac{ \pi \hbar \sqrt{2 \pi p}}{ae}\right).
\label{eq:3}
\end{equation}
Note that in Fig. 2(b) our deduced values of $B_E$ are indeed close to the maximum of the positive magnetoresistance peak often used as a measure of the extinction field.

In closing, we remark on the following observations. First, there is a sample dependence of $ \eta$, evident from the two Hall bars of the $a=175$ nm sample, where $ \eta$ is more than a factor of 2 different. At present, we do not know what causes this sample dependence but we suspect it is related to the details of the patterning process. Second, the values of $ \eta$ are comparable to those reported for electron systems \cite{ref10Endo.PRB.62.16761.2000} with similar parameters \cite{ref19footnote5} although a direct comparison is difficult since $ \eta$ depends on the modulation period, the QW depth from the sample surface, the carrier density, and the effective mass.

\vspace{5 mm}

\begin{acknowledgements}

We are grateful to A. Endo and Y. Iye for useful discussions. We also thank Tokoyama Corporation for supplying the negative e-beam resist TEBN-1 used to make the samples. We acknowledge support through the NSF (ECCS-1001719, DMR-0904117, and MRSEC DMR-0819860) and the Moore Foundation for sample fabrication and characterization, and the DOE BES (DE-FG02-00-ER45841) for measurements.
\end{acknowledgements}.

\end{document}